\documentclass[cameraready]{Interspeech}
\title{Progressive Alignment Objectives for Aligner-Encoder based ASR}

\author[affiliation={1}]{Jaeyoung}{Lee}
\author[affiliation={1}]{Masato}{Mimura}
\author[affiliation={1}]{Takafumi}{Moriya}
\address{
    $^1$ NTT, Inc., Japan
}

\email{jae.lee@ntt.com}


\usepackage{comment}
\usepackage{multirow}
\usepackage{bm}

\begin{document}

\maketitle

\begin{abstract}
Aligner-Encoders are recently proposed seq2seq end-to-end ASR models that replace decoder attention by predicting the u-th token directly from the u-th encoder position, so the encoder must learn the alignment internally without cross-attention or a transducer lattice.
In practice, this alignment often forms abruptly in the upper layers, making training sensitive and brittle on long utterances.
We propose InterAligner, which adds an intermediate Aligner objective so alignment can form progressively across depth, together with an intermediate CTC loss (InterCTC) to stabilize optimization. On LibriSpeech with a 17-layer Conformer, a final-only Aligner reaches 5.0/7.8 WER (test-clean/other). InterCTC improves to 3.4/6.0, and InterAligner further reduces WER to 3.1/5.6 with the largest gains on long utterances.
\end{abstract}

\section{Introduction}
\label{sec:intro}

Learning a monotonic alignment between acoustic frames and output tokens remains central to end-to-end (E2E) automatic speech recognition (ASR). Classical E2E formulations differ primarily in how they represent alignment: Connectionist Temporal Classification (CTC) \cite{graves2006ctc} marginalizes over monotonic paths with dynamic programming, RNN-Transducer (RNN-T) \cite{graves2012rnnt} extends this idea with a joint acoustic-label lattice, and attention-based encoder-decoders (AED) learn alignments implicitly through cross-attention \cite{chan2016las}. While modern encoders such as the Conformer \cite{gulati20_interspeech} have strengthened all three paradigms, alignment learning can still dominate optimization difficulty and robustness, especially for long utterances.

Beyond these established paradigms, some recent sequence-to-sequence (seq2seq) ASR frameworks aim to reduce reliance on implicit, unconstrained cross-attention alignments by introducing monotonic inductive bias or structured emission mechanisms. Hybrid objectives that combine AED training with auxiliary monotonic losses such as CTC can improve stability and convergence \cite{kim2017jointctcattn}.
Separately, Continuous Integrate-and-Fire (CIF)~\cite{dong2020cif} introduces a soft monotonic emission process that aggregates frame-level representations into token-level segments by dynamically ``integrating'' evidence until a threshold triggers output. Collectively, these works illustrate complementary ways of injecting monotonic bias via auxiliary losses or explicit emission mechanisms, and suggest that architectures where alignment must form within the network (e.g., within encoder self-attention) may benefit from additional monotonic guidance at intermediate depth.

Aligner-Encoders \cite{stooke2024aligner} offer a recent alternative that makes alignment explicit inside encoder self-attention, enabling lightweight decoding while remaining competitive with strong AED baselines.
However, Aligner-Encoders exhibit a distinctive training dynamic: clear diagonal alignment emerges abruptly only in a small number of top layers of a 17-layer encoder \cite{stooke2024aligner}.
This creates a late-layer alignment bottleneck, where the model must transition from largely unaligned acoustic representations to near-monotonic alignment within just a few layers.
In particular, for long utterances, a large mismatch between encoder-frame counts and token length makes alignment difficult and markedly degrades recognition performance \cite{stooke2024aligner}.

We propose \textbf{InterAligner}, which encourages alignment to form progressively across depth by adding an intermediate Aligner objective at an upper-intermediate layer using a longer, finer-grained target sequence, followed by the final Aligner objective at the last layer with a shorter, coarser tokenization.
We additionally attach an early-layer intermediate CTC loss (InterCTC), building on intermediate CTC supervision \cite{lee2021interctc} and CTC-assisted training \cite{kim2017jointctcattn}.
Our method creates a curriculum over alignment difficulty and is related to hierarchical/multi-granularity supervision \cite{sanabria2018slt,higuchi2022icassp,kusunoki24_interspeech}, but specifically targets alignment emergence within encoder self-attention.

This work (i) establishes InterCTC as an effective optimization aid for Aligner-Encoders and (ii) introduces InterAligner to encourage progressive alignment formation across depth.
We show consistent accuracy gains from both components on LibriSpeech and Common Voice English, and provide complementary evidence through long-utterance analysis and attention visualizations.

\begin{figure*}[t]
  \centering
  \includegraphics[width=\textwidth]{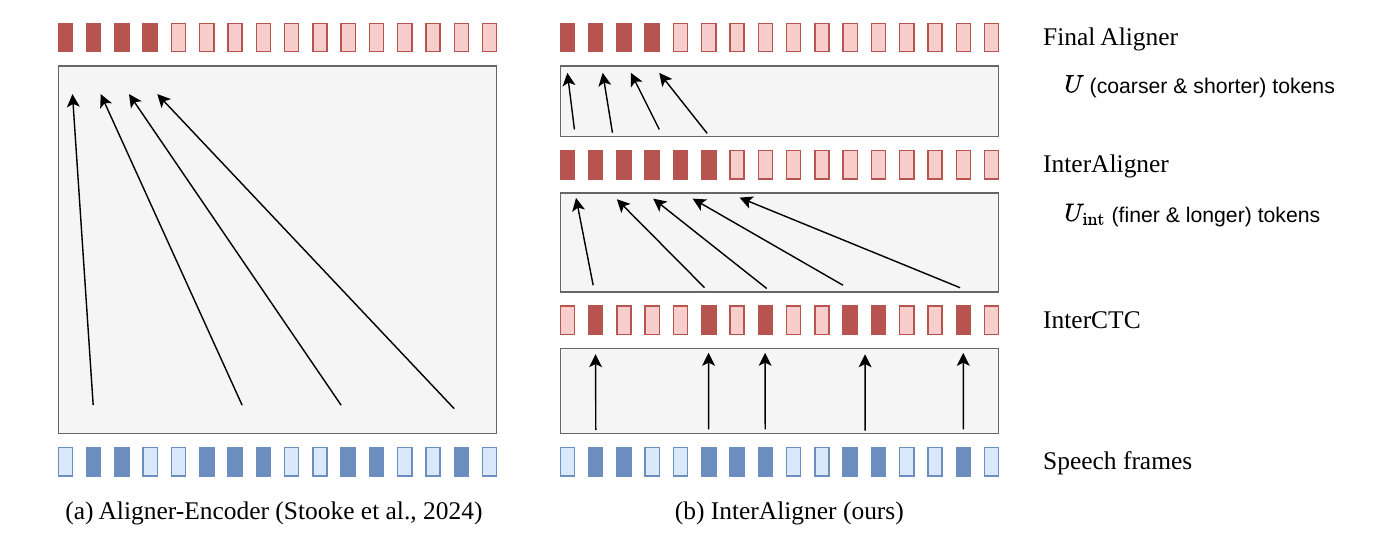}
  \vspace{-6mm}
  \caption{InterAligner augments Aligner-Encoders with auxiliary monotonic supervision at intermediate depth, by providing early monotonic guidance on a longer sequence with a finer token granularity.}
  \label{fig:interaligner}
\end{figure*}

\begin{figure}[t]
  \centering
  \vspace{-8mm}
  \includegraphics[width=\columnwidth]{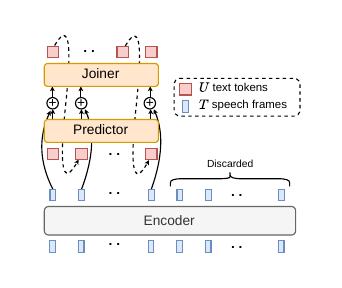}
  \vspace{-10mm}
  \caption{Aligner-encoder architecture. The joiner and predictor modules operate autoregressively on $U$ speech frames; speech after that is discarded.}
  \label{fig:aligner-arch}
\end{figure}

\section{Related Work}
\label{sec:related}

\subsection{Aligner-Encoders and alignment-explicit transduction}
E2E ASR research has repeatedly revisited the question of how to represent and learn monotonic audio-text alignment reliably \cite{prabhavalkar2024e2easrsurvey}.
Beyond CTC and RNN-T, several alignment-explicit transduction frameworks have been proposed to reduce reliance on unconstrained soft attention, including the Recurrent Neural Aligner (RNA) \cite{sak17_interspeech} and self-attention based transducer variants \cite{tian19b_interspeech,yeh2019transformertransducer,zhang2020transformertransducer}.
Relatedly, self-attention can be augmented with an explicit alignment mechanism for latency-controlled recognition \cite{dong2019selfattentionaligner}, and label-synchronous transducers further emphasize label-level (not frame-level) representation and emission \cite{deng2023lstransducer}.

Stooke et al.\ proposed \emph{Aligner-Encoders}, which expose audio-text alignment within encoder self-attention and enable lightweight decoding while remaining competitive with attention-based encoder-decoders (AED) under a shared Conformer backbone \cite{stooke2024aligner}.
A key observation in Aligner-Encoders is that clear diagonal alignment structure typically becomes prominent only in the upper layers of deep encoders \cite{stooke2024aligner}.
Our work builds directly on this setting and addresses the resulting depth bottleneck by distributing alignment supervision across depth so that alignment can form more gradually, improving robustness particularly for long utterances.

\subsection{Monotonic bias and intermediate supervision}
Monotonicity is commonly encouraged via auxiliary CTC objectives in AED training \cite{kim2017jointctcattn,watanabe2017hybridctcattention} and joint decoding \cite{hori2017jointctcattentiondecoding}. Structured monotonic attention and related mechanisms further constrain alignment for streaming/robustness \cite{raffel2017monotonic,chiu2018mocha,zhao20j_interspeech,moritz2019triggeredattention,dong2020cif}, motivated by analyses of attention alignment behavior \cite{prabhavalkar17b_interspeech}. Intermediate/deep supervision improves optimization in deep encoders \cite{lee2021interctc,tjandra2020dejavu}; we evaluate InterCTC in Aligner-Encoders and introduce an intermediate Aligner objective to directly guide self-attention alignment formation.

\subsection{Hierarchical and multi-granularity supervision}
Hierarchical supervision attaches losses at multiple depths and/or label granularities \cite{sanabria2018slt,krishna2018hierarchicalmultitaskctc,higuchi2022icassp,kusunoki24_interspeech,moriya18_interspeech}. 
InterAligner is closest in spirit to this line of hierarchical/multi-granularity supervision, but is tailored to Aligner-Encoders: we apply an intermediate \emph{Aligner} loss on a longer, finer-grained target before the final Aligner loss, encouraging monotonic alignment to emerge progressively within encoder self-attention.

\section{Methodology}
\label{sec:method}

\subsection{Aligner-Encoders}

We follow the Aligner-Encoder formulation of \cite{stooke2024aligner} (see Figure~\ref{fig:aligner-arch}).
Let $\bm{X}=(\bm{x}_1,\dots,\bm{x}_T)$ denote an input acoustic feature sequence of length $T$.
Let $\bm{y}=(y_1,\dots,y_U)$ be the output token sequence of length $U$, where we append an end-of-sequence token $\langle\mathrm{eos}\rangle$ (and include it in $U$).
Aligner-Encoders assume $U \le T'$, where $T'$ is the encoder output length after subsampling.

An encoder network \(f_{\mathrm{enc}}\) maps $\bm{X}$ to an acoustic embedding sequence
\begin{equation}
\bm{H} = f_{\mathrm{enc}}(\bm{X}) = (\bm{h}_1,\dots,\bm{h}_{T'}),
\label{eq:enc}
\end{equation}
and a prediction network \(f_{\mathrm{pred}}\) produces a text embedding sequence conditioned on the token history:
\begin{equation}
\bm{g}_u = f_{\mathrm{pred}}(\bm{g}_{u-1}, y_{u-1}), \qquad u=1,\dots,U.
\label{eq:pred}
\end{equation}
We set $y_0=\langle\mathrm{sos}\rangle$ and initialize $\bm{g}_0$ (e.g., all zeros). In our implementation, $f_{\mathrm{pred}}$ is a one-layer LSTM over token embeddings.

A joiner network \(f_{\mathrm{joint}}\) combines the paired acoustic/text embeddings to produce logits over the vocabulary:
\begin{align}
\bm{z}_u &= f_{\mathrm{joint}}(\bm{h}_u, \bm{g}_u), \\
P(y_u \mid \bm{X}, y_{<u}) &= \mathrm{softmax}(\bm{z}_u).
\label{eq:joint}
\end{align}
Throughout the paper, $f_{\mathrm{joint}}$ is a feed-forward joiner:
\begin{equation}
\tilde{\bm{z}}_u = \tanh(\bm{W}_{h}\bm{h}_u + \bm{W}_{g}\bm{g}_u + \bm{b}), \qquad
\bm{z}_u = \bm{W}_{o}\tilde{\bm{z}}_u + \bm{b}_{o},
\label{eq:joiner_mlp}
\end{equation}
followed by $\mathrm{softmax}(\cdot)$ in Eq.~\eqref{eq:joint}.
The defining restriction of Aligner-Encoders is the one-to-one pairing $(\bm{h}_u, \bm{g}_u)$ in Eq.~\eqref{eq:joint}, rather than a full $T'\times U$ lattice as in RNN-T.

Training minimizes the negative log-likelihood of the label sequence:
\begin{equation}
\mathcal{L}_{\mathrm{final}}(\theta)
= - \sum_{u=1}^{U} \log P(y_u \mid \bm{X}, y_{<u}; \theta).
\label{eq:align_loss}
\end{equation}
With $T' \ge U$, the loss is applied only up to $u=U$ and the remaining encoder frames are ignored, requiring the encoder to move label-relevant information into its first $U$ outputs.

At inference time, given $\bm{h}_1,\dots,\bm{h}_{T'}$, decoding proceeds sequentially over encoder frames and emits one token per frame until $\langle\mathrm{eos}\rangle$ is produced, as in standard AED.
Let $\hat y_u$ denote the token emitted at step $u$.
Starting from $\hat y_0=\langle\mathrm{sos}\rangle$ and state $\bm{g}_0$, for $u=1,2,\dots,T'$ we compute
\begin{align}
\bm{g}_u &= f_{\mathrm{pred}}(\bm{g}_{u-1}, \hat y_{u-1}),\\
\bm{p}_u(\cdot) &= \mathrm{softmax}(f_{\mathrm{joint}}(\bm{h}_u,\bm{g}_u)),
\label{eq:decode_step}
\end{align}
where we select $\hat y_u$ from $\bm{p}_u(\cdot)$ using greedy decoding or beam search, and stop when $\hat y_u=\langle\mathrm{eos}\rangle$.

\begin{table}[t]
\centering
\caption{Main LibriSpeech results (WERs \%) on test sets.}
%Last row, \emph{concat}, employs two-utterance concatenation during training, filtering $>35\,\mathrm{s}$.
\label{tab:main_ls}
\begin{tabular}{lcc}
\hline
System & test-clean & test-other \\
\hline
Final Aligner only (Stooke et al.) & 4.8 & 6.5 \\
Final Aligner only (ours) & 5.0 & 7.8 \\
\quad+ InterCTC & 3.4 & 6.0 \\
\quad\quad+ InterAligner & 3.1 & 5.6 \\
% \quad\quad\quad+ concat & 2.5 & 5.3 \\
\hline
\end{tabular}
\end{table}

\begin{table}[t]
\centering
\caption{Common Voice (16.1) English results (WERs \%).}
\label{tab:cv_en}
\begin{tabular}{lc}
\hline
System & test \\
\hline
Final Aligner only & 12.4 \\
\quad+ InterCTC & 11.2 \\
\quad\quad+ InterAligner & 10.9 \\
\hline
\end{tabular}
\end{table}

\subsection{InterAligner and InterCTC}
InterAligner augments the baseline Aligner-Encoder objective with two intermediate losses, as illustrated in Fig.~\ref{fig:interaligner}(b).
Let $\bm{H}^{(\ell)}=(\bm{h}^{(\ell)}_1,\dots,\bm{h}^{(\ell)}_{T'})$ denote the encoder sequence after layer $\ell$ ($\ell=1,\dots,L$), with $\bm{H}^{(L)}\equiv \bm{H}$.

\begin{table}[t]
\centering
\caption{WERs (\%) by utterance duration on LibriSpeech test.}
\label{tab:length}
\setlength{\tabcolsep}{5pt} % default ~6pt
\begin{tabular}{llcccc}
\hline
Split & System & $<$17s & 17--21s & $>$21s & All \\
\hline
\multirow{3}{*}{clean} &
Final Aligner only & 3.2 & 5.7 & 23.4 & 5.0 \\
& \quad + InterCTC & 2.3 & 2.3 & 17.0 & 3.4 \\
& \quad\quad + InterAligner & 2.4 & 2.9 & 11.6 & 3.1 \\
\hline
\multirow{3}{*}{other} &
Final Aligner only & 7.0 & 8.2 & 24.0 & 7.8 \\
& \quad + InterCTC & 5.4 & 5.3 & 18.0 & 6.0 \\
& \quad\quad + InterAligner & 5.2 & 5.5 & 13.5 & 5.6 \\
\hline
\end{tabular}
\end{table}

InterCTC attaches an intermediate CTC loss $\mathcal{L}_{\mathrm{ctc}}$ at layer $\ell_{\mathrm{ctc}}$ (we use $\ell_{\mathrm{ctc}}=12$).
InterAligner attaches an intermediate Aligner loss at layer $\ell_{\mathrm{int}}$ (we use $\ell_{\mathrm{int}}=15$ in the main results) using an intermediate token sequence
\begin{equation}
\bm{y}^{\mathrm{int}}=(y^{\mathrm{int}}_1,\dots,y^{\mathrm{int}}_{U_{\mathrm{int}}}),
\label{eq:int_targets}
\end{equation}
derived from the same transcript with a finer tokenization, typically yielding $U_{\mathrm{int}} > U$.
The InterAligner head uses a separate predictor and joiner from the final Aligner (no parameter sharing).
We ensure $U_{\mathrm{int}}\le T'$ so that the one-to-one pairing in Aligner-Encoders is well-defined.
Using the same predictor–joiner factorization as Eqs.~\eqref{eq:pred}–\eqref{eq:joint}, but applied to $\bm{H}^{(\ell_{\mathrm{int}})}$ and $\bm{y}^{\mathrm{int}}_{<u}$, the intermediate Aligner loss is
\begin{equation}
\mathcal{L}_{\mathrm{int}}(\theta)
= - \sum_{u=1}^{U_{\mathrm{int}}} \log P_{\mathrm{int}}\left(y^{\mathrm{int}}_u \mid \bm{X}, y^{\mathrm{int}}_{<u}; \theta\right).
\label{eq:int_align_loss}
\end{equation}

These objectives are intended to shape alignment formation in stages: InterCTC encourages token-predictive representations early in the encoder, InterAligner encourages alignment to a longer, finer-grained sequence at intermediate depth, and the final Aligner loss $\mathcal{L}_{\mathrm{final}}$ (Eq.~\eqref{eq:align_loss}) encourages refinement to the shorter, coarser target at the top layer.

The overall training objective is
\begin{equation}
\mathcal{L}(\theta)
= \lambda_{\mathrm{final}}\mathcal{L}_{\mathrm{final}}(\theta)
+ \lambda_{\mathrm{int}}\,\mathcal{L}_{\mathrm{int}}(\theta)
+ \lambda_{\mathrm{ctc}}\,\mathcal{L}_{\mathrm{ctc}}(\theta).
\label{eq:total_loss}
\end{equation}

\section{Experimental Evaluations}
\label{sec:exp}
\subsection{Models and Datasets}

We evaluate on LibriSpeech 960h and report WER on test-clean and test-other.
We additionally evaluate on Common Voice 16.1 English, where punctuation is removed in both train and test.

All systems use the same 17-layer Conformer-L encoder (overall model $\sim$118M parameters).
Unless otherwise noted, the final Aligner head is attached at layer 17, InterAligner at layer 15, and InterCTC at layer 12.
For a fair comparison with \cite{stooke2024aligner}, we set the beam width to 6.
For LibriSpeech, we train for 100 epochs and report results with model averaging over the 10 best checkpoints.
For Common Voice, we train for 50 epochs with the same 10-best averaging.
The effective batch size is set to \(\sim\)2 hours of audio in all settings.
For reference, Table~\ref{tab:main_ls} includes the final-only Aligner WER reported in Stooke et al. \cite{stooke2024aligner}. 
Reproducing that number is challenging under a limited training budget, which motivates adding intermediate objectives (InterCTC, InterAligner) to stabilize training.

We use standard Transformer warmup/decay with 20k warmup steps.
Peak learning rate is found optimal when 0.0020 for settings that do not use any target vocabulary size $\le256$, and 0.0025 otherwise.
The CTC weight is fixed to $\lambda_{\mathrm{ctc}}=0.1$ in all experiments.
For InterAligner systems, we tune $(\lambda_{\mathrm{final}},\lambda_{\mathrm{int}})$ and report the best-performing weights; ablations include explicit values.

We compare progressively augmented systems: a final-only Aligner, the same model with InterCTC, and the full InterAligner configuration (InterCTC + InterAligner). All tokenization uses Byte Pair Encoding (BPE) \cite{sennrich2016subword}. Unless stated otherwise, the number of tokens in the vocabulary for the final Aligner is 1024, InterAligner uses 256, and InterCTC matches the tokenization of the Aligner objective at the nearest higher layer (1024 in the InterCTC setting; 256 in InterAligner).

% abstract too abstract, first sentence too long

\subsection{Results}
Table~\ref{tab:main_ls} reports the main LibriSpeech results.
InterCTC yields a large improvement over the final-only Aligner baseline, and adding InterAligner provides a further gain.
% We also train a variant with concatenated training utterances to reduce the mismatch between training and long test utterances.
% Specifically, we concatenate LibriSpeech samples twice during training and filter utterances longer than 35s.
% This setting further improves the same InterAligner configuration.
We observe the same trend on Common Voice English (Table~\ref{tab:cv_en}) as on LibriSpeech: adding InterCTC improves over the final-only Aligner, and adding InterAligner yields further gains. 
Final-only Aligner obtains 12.4\% WER, adding InterCTC improves to 11.2\%, and adding InterAligner further reduces WER to 10.9\% on the final head.
We also verified with statistical significance testing that the WER reductions from InterCTC and from adding InterAligner are significant at the 1\% level.

Table~\ref{tab:length} reports WER stratified by utterance duration.
While the InterAligner system matches the InterCTC baseline on short and medium segments, it substantially improves performance on long utterances ($>21$ s), reducing WER from 17.0 to 11.6 on test-clean and from 18.0 to 13.5 on test-other.
These gains support our motivation that distributing alignment supervision across depth makes Aligner-Encoders more robust when the model must learn alignment over longer contexts.

We next analyze ablations that isolate the roles of intermediate tokenization, hierarchical supervision, and auxiliary-objective design.
Table~\ref{tab:abl_match_weight} evaluates whether the intermediate CTC target should match the InterAligner target, and how sensitive performance is to the loss weights, with $\lambda_{\mathrm{ctc}}=0.1$ for all rows.
Using the same intermediate vocabulary size for InterAligner and InterCTC consistently outperforms using a larger InterCTC vocabulary (i.e., matching 256/256 vs.\ mismatching 256/1024).
The results also show that the relative weighting of the final and intermediate Aligner losses matters: emphasizing the intermediate Aligner loss $(\lambda_{\mathrm{final}},\lambda_{\mathrm{int}})=(0.5,1.0)$ is best in the matched setting, while $(1.0,0.5)$ degrades the intermediate head and slightly worsens the final head.
Interestingly, under the optimal weighting the intermediate head attains slightly lower WER than the final head (3.0/5.5 vs.\ 3.1/5.6).
All main results are reported using decoding from the final head.

\begin{table}[t]
\centering
\caption{Effect of matching InterAligner/InterCTC targets and loss-weight sensitivity. Left: target BPE vocabulary sizes for the final Aligner (final), InterAligner (inter), and InterCTC.
Right: WERs (\%) on LibriSpeech test sets.}
\label{tab:abl_match_weight}
\begin{tabular}{ccc|c|cc}
\hline
\multicolumn{3}{c|}{Target BPE} & Weights & \multicolumn{2}{c}{test clean / other} \\
final & inter & CTC & $\lambda_{\mathrm{final}} / \lambda_{\mathrm{int}}$ & final & inter \\
\hline
1024 & 256  & 256  & 0.5 / 1.0 & \textbf{3.1} / \textbf{5.6} & \textbf{3.0} / \textbf{5.5} \\
1024 & 256  & 256  & 1.0 / 0.5 & \textbf{3.1} / 5.8 & 5.7 / 7.0 \\
1024 & 256  & 1024 & 1.0 / 0.5 & 3.2 / 5.8 & 4.0 / 6.1 \\
\hline
\end{tabular}
\end{table}

\begin{table}[t]
\centering
\caption{Ablation results on tokenization choices. Left: target BPE vocabulary sizes.
Right: WERs (\%) on LibriSpeech test sets.
``--'' denotes that the corresponding objective is not used.}
\label{tab:abl_misc}
\begin{tabular}{ccc|cc}
\hline
\multicolumn{3}{c|}{Target BPE} & \multicolumn{2}{c}{test clean / other} \\
final & inter & CTC & final & inter \\
\hline
1024 & 1024 & 1024 & 3.7 / 6.3 & 3.8 / 6.3 \\
1024 & 256  & 256  & 3.1 / 5.8 & 5.7 / 7.0 \\
1024 & 64   & 64   & \textbf{3.0} / \textbf{5.7} & 5.7 / 6.8 \\
256  & --   & 256  & 5.0 / 6.4 & -- \\
\hline
\end{tabular}
\end{table}

Table~\ref{tab:abl_misc} varies the target vocabulary sizes used for the intermediate objectives, with 
$(\lambda_{\mathrm{final}},\lambda_{\mathrm{int}},\lambda_{\mathrm{ctc}}) = (1.0, 0.5, 0.1)$, except for the last row, where $(\lambda_{\mathrm{final}},\lambda_{\mathrm{int}},\lambda_{\mathrm{ctc}}) = (1.0, 0, 0.1)$.
Using smaller intermediate vocabularies (256 or 64) yields strong performance, whereas using the same intermediate vocabulary as the final Aligner (1024) degrades WER, indicating that intermediate granularity substantially affects the difficulty of the implicit alignment problem.
The results also indicate that reducing vocabulary size alone is insufficient: removing InterAligner and training only with 256 (final=256, CTC=256) performs much worse than the hierarchical configuration, suggesting that the gains arise from progressive supervision rather than tokenization alone.

\begin{figure}[t]
  \centering
  \includegraphics[width=\columnwidth]{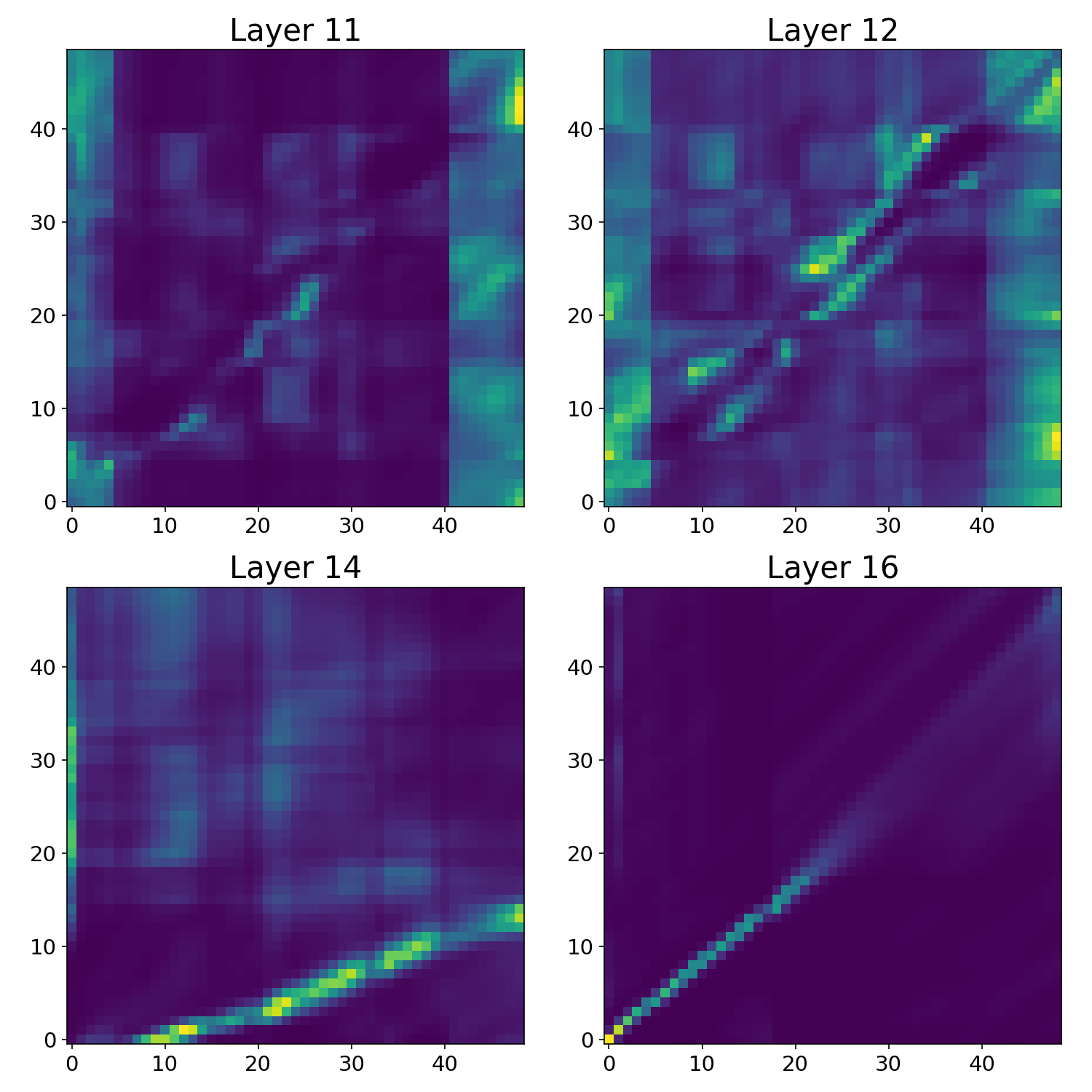}
  \caption{Averaged (8-head) encoder self-attention maps for InterAligner on a LibriSpeech utterance ($U_{\mathrm{int}}=11$, $U=9$), illustrating progressive alignment emergence across layers ($T' \rightarrow U_{\mathrm{int}}$ in layer 14, $U_{\mathrm{int}} \rightarrow U$ in layer 16).}
  \label{fig:attn}
\end{figure}

\begin{table}[t]
\centering
\caption{Effect of InterAligner layer placement for InterAligner+InterCTC, reporting WER (\%) on test clean / other.}
\label{tab:abl_layer}
\begin{tabular}{ccc}
\hline
InterAligner layer & final & inter \\
\hline
16th & 3.8 / 5.9 & 3.5 / 5.8 \\
15th & \textbf{3.1} / \textbf{5.6} & \textbf{3.0} / \textbf{5.5} \\
13th & 3.5 / 6.4 & 3.3 / 5.9 \\
\hline
\end{tabular}
\end{table}

Table~\ref{tab:abl_layer} examines the placement of the InterAligner objective, with $(\lambda_{\mathrm{final}},\lambda_{\mathrm{int}},\lambda_{\mathrm{ctc}})=(0.5,1.0,0.1)$ and an intermediate vocabulary size of 256 for both InterAligner and InterCTC.
The best results are obtained at layer 15, whereas attaching InterAligner at layer 16 substantially degrades performance. One interpretation is that at least two layers are needed to convert the intermediate alignment into the final alignment, consistent with the findings in \cite{stooke2024aligner} where alignment is reported to emerge across 2 upper layers.

\subsection{Attention Visualization}
\vspace{-1mm}
Figure~\ref{fig:attn} shows averaged (8-head) encoder self-attention maps for a representative LibriSpeech utterance ($U_{\mathrm{int}}=11$, $U=9$). Alignment does not emerge in the lower layers shown (11-12), but appears progressively in the upper encoder layers.
At layer 14, attention concentrates into a sharp monotonic band consistent with $T' \rightarrow U_{\mathrm{int}}$ alignment. By layer 16, a second sharp diagonal pattern is observed at the coarser final granularity, consistent with $U_{\mathrm{int}} \rightarrow U$. These patterns support our claim that InterAligner encourages staged alignment formation across depth rather than direct $T' \rightarrow U$ alignment in the top layers.

% TODO explain (U_int -> U) alignment pattern continues beyond (11,9)

\vspace{-1mm}
\section{Conclusion}
\label{sec:conclusion}

We proposed InterAligner, which augments Aligner-Encoders with progressive alignment supervision by adding an intermediate Aligner objective, together with an intermediate CTC loss (InterCTC) for optimization stability.
Across LibriSpeech and Common Voice English, InterAligner consistently improves over a final-only Aligner and over InterCTC alone, with the largest gains on long utterances.
Attention visualizations support the intended emergence of progressive alignment across depth.
Future work will study progressive objectives for streaming and long-form recognition and their interaction with language-model integration.

\section{Generative AI Use Disclosure}
We used generative AI tools to assist with code development and manuscript editing; all generated content was reviewed and validated by the authors.

\bibliographystyle{IEEEtran}
\bibliography{mybib}

@inproceedings{moriya18_interspeech,
  title     = {{Multi-task Learning with Augmentation Strategy for Acoustic-to-word Attention-based Encoder-decoder Speech Recognition}},
  author    = {Takafumi Moriya and Sei Ueno and Yusuke Shinohara and Marc Delcroix and Yoshikazu Yamaguchi and Yushi Aono},
  year      = {2018},
  booktitle = {{Interspeech 2018}},
  pages     = {2399--2403},
  doi       = {10.21437/Interspeech.2018-1866},
  issn      = {2958-1796},
}

@inproceedings{sennrich2016subword,
  author    = {Rico Sennrich and Barry Haddow and Alexandra Birch},
  title     = {Neural Machine Translation of Rare Words with Subword Units},
  booktitle = {Proceedings of the 54th Annual Meeting of the Association for Computational Linguistics (Volume 1: Long Papers)},
  pages     = {1715--1725},
  year      = {2016},
  doi       = {10.18653/v1/P16-1162}
}

@inproceedings{sak17_interspeech,
  title     = {{Recurrent Neural Aligner: An Encoder-Decoder Neural Network Model for Sequence to Sequence Mapping}},
  author    = {Hasim Sak and Kanishka Rao and Fran{\c{c}}oise Beaufays},
  year      = {2017},
  booktitle = {{Interspeech 2017}},
  pages     = {1298--1302},
  doi       = {10.21437/Interspeech.2017-149}
}

@inproceedings{dong2019selfattentionaligner,
  author    = {Linhao Dong and Feng Wang and Bo Xu},
  title     = {Self-Attention Aligner: A Latency-Controlled Self-Attention Model for Automatic Speech Recognition},
  booktitle = {2019 IEEE International Conference on Acoustics, Speech and Signal Processing (ICASSP)},
  pages     = {5656--5660},
  year      = {2019},
  doi       = {10.1109/ICASSP.2019.8682954}
}

@inproceedings{raffel2017monotonic,
  author    = {Colin Raffel and Minh-Thang Luong and Peter J. Liu and Ron J. Weiss and Douglas Eck},
  title     = {Online and Linear-Time Attention by Enforcing Monotonic Alignments},
  booktitle = {Proceedings of the 34th International Conference on Machine Learning (ICML)},
  pages     = {2837--2846},
  year      = {2017}
}

@inproceedings{chiu2018mocha,
  author    = {Chung-Cheng Chiu and Colin Raffel},
  title     = {Monotonic Chunkwise Attention},
  booktitle = {International Conference on Learning Representations (ICLR)},
  year      = {2018}
}

@inproceedings{moritz2019triggeredattention,
  author    = {Niko Moritz and Takaaki Hori and Jonathan Le Roux},
  title     = {Triggered Attention for End-to-End Speech Recognition},
  booktitle = {2019 IEEE International Conference on Acoustics, Speech and Signal Processing (ICASSP)},
  pages     = {5666--5670},
  year      = {2019},
  doi       = {10.1109/ICASSP.2019.8683510}
}

@inproceedings{zhang2020transformertransducer,
  author    = {Qian Zhang and Han Lu and Hasim Sak and Anshuman Tripathi and Erik McDermott and Stephen Koo and Shankar Kumar},
  title     = {Transformer Transducer: A Streamable Speech Recognition Model with Transformer Encoders and RNN-T Loss},
  booktitle = {2020 IEEE International Conference on Acoustics, Speech and Signal Processing (ICASSP)},
  pages     = {7829--7833},
  year      = {2020},
  doi       = {10.1109/ICASSP40776.2020.9053896}
}

@article{yeh2019transformertransducer,
  author  = {Ching-Feng Yeh and Jay Mahadeokar and Kaustubh Kalgaonkar and Yongqiang Wang and Duc Le and Mahaveer Jain and Kjell Schubert and Christian Fuegen and Michael L. Seltzer},
  title   = {Transformer-Transducer: End-to-End Speech Recognition with Self-Attention},
  journal = {arXiv preprint arXiv:1910.12977},
  year    = {2019}
}

@inproceedings{tian19b_interspeech,
  title     = {{Self-Attention Transducers for End-to-End Speech Recognition}},
  author    = {Zhengkun Tian and Jiangyan Yi and Jianhua Tao and Ye Bai and Zhengqi Wen},
  year      = {2019},
  booktitle = {{Interspeech 2019}},
  pages     = {4395--4399},
  doi       = {10.21437/Interspeech.2019-2203}
}

@ARTICLE{deng2023lstransducer,
  author={Deng, Keqi and Woodland, Philip C.},
  journal={IEEE/ACM Transactions on Audio, Speech, and Language Processing}, 
  title={Label-Synchronous Neural Transducer for Adaptable Online E2E Speech Recognition}, 
  year={2024},
  volume={32},
  number={},
  pages={3507-3516},
  doi={10.1109/TASLP.2024.3419421}}

@inproceedings{prabhavalkar17b_interspeech,
  title     = {{An Analysis of {``Attention''} in Sequence-to-Sequence Models}},
  author    = {Rohit Prabhavalkar and Tara N. Sainath and Bo Li and Kanishka Rao and Navdeep Jaitly},
  year      = {2017},
  booktitle = {{Interspeech 2017}},
  pages     = {3702--3706},
  doi       = {10.21437/Interspeech.2017-232}
}

@article{watanabe2017hybridctcattention,
  author  = {Shinji Watanabe and Takaaki Hori and Suyoun Kim and John R. Hershey and Tomoki Hayashi},
  title   = {Hybrid {CTC}/Attention Architecture for End-to-End Speech Recognition},
  journal = {IEEE Journal of Selected Topics in Signal Processing},
  volume  = {11},
  number  = {8},
  pages   = {1240--1253},
  year    = {2017},
  doi     = {10.1109/JSTSP.2017.2763455}
}

@inproceedings{hori2017jointctcattentiondecoding,
  author    = {Takaaki Hori and Shinji Watanabe and John Hershey},
  title     = {Joint {CTC}/attention decoding for end-to-end speech recognition},
  booktitle = {Proceedings of the 55th Annual Meeting of the Association for Computational Linguistics (Volume 1: Long Papers)},
  pages     = {518--529},
  year      = {2017},
  doi       = {10.18653/v1/P17-1048}
}

@inproceedings{tjandra2020dejavu,
  author    = {Andros Tjandra and Chunxi Liu and Frank Zhang and Xiaohui Zhang and Yongqiang Wang and Gabriel Synnaeve and Satoshi Nakamura and Geoffrey Zweig},
  title     = {Deja-vu: Double Feature Presentation and Iterated Loss in Deep Transformer Networks},
  booktitle = {2020 IEEE International Conference on Acoustics, Speech and Signal Processing (ICASSP)},
  pages     = {6899--6903},
  year      = {2020},
  doi       = {10.1109/ICASSP40776.2020.9052964}
}

@article{krishna2018hierarchicalmultitaskctc,
  author  = {Kalpesh Krishna and Shubham Toshniwal and Karen Livescu},
  title   = {Hierarchical Multitask Learning for {CTC}-based Speech Recognition},
  journal = {arXiv preprint arXiv:1807.06234},
  year    = {2018}
}

@inproceedings{zhao20j_interspeech,
  title     = {{Cross Attention with Monotonic Alignment for Speech Transformer}},
  author    = {Yingzhu Zhao and Chongjia Ni and Cheung-Chi Leung and Shafiq Joty and Eng Siong Chng and Bin Ma},
  year      = {2020},
  booktitle = {{Interspeech 2020}},
  pages     = {5031--5035},
  doi       = {10.21437/Interspeech.2020-1198}
}

@article{prabhavalkar2024e2easrsurvey,
  author  = {Rohit Prabhavalkar and Takaaki Hori and Tara N. Sainath and Ralf Schl{\"u}ter and Shinji Watanabe},
  title   = {End-to-End Speech Recognition: A Survey},
  journal = {IEEE/ACM Transactions on Audio, Speech, and Language Processing},
  volume  = {32},
  pages   = {325--351},
  year    = {2024},
  doi     = {10.1109/TASLP.2023.3328283}
}

@inproceedings{graves2006ctc,
  author    = {Alex Graves and Santiago Fern{\'a}ndez and Faustino Gomez and J{\"u}rgen Schmidhuber},
  title     = {Connectionist Temporal Classification: Labelling Unsegmented Sequence Data with Recurrent Neural Networks},
  booktitle = {Proceedings of the 23rd International Conference on Machine Learning (ICML)},
  pages     = {369--376},
  year      = {2006},
  doi       = {10.1145/1143844.1143891}
}

@article{graves2012rnnt,
  author  = {Alex Graves},
  title   = {Sequence Transduction with Recurrent Neural Networks},
  journal = {arXiv preprint arXiv:1211.3711},
  year    = {2012}
}

@inproceedings{chan2016las,
  author    = {William Chan and Navdeep Jaitly and Quoc V. Le and Oriol Vinyals},
  title     = {Listen, Attend and Spell: A Neural Network for Large Vocabulary Conversational Speech Recognition},
  booktitle = {2016 IEEE International Conference on Acoustics, Speech and Signal Processing (ICASSP)},
  pages     = {4960--4964},
  year      = {2016},
  doi       = {10.1109/ICASSP.2016.7472621}
}

@inproceedings{gulati20_interspeech,
  title     = {{Conformer: Convolution-augmented Transformer for Speech Recognition}},
  author    = {Anmol Gulati and James Qin and Chung-Cheng Chiu and Niki Parmar and Yu Zhang and Jiahui Yu and Wei Han and Shibo Wang and Zhengdong Zhang and Yonghui Wu and Ruoming Pang},
  year      = {2020},
  booktitle = {Proc. {INTERSPEECH} 2020 -- 21\textsuperscript{st} Annual Conference of the International Speech Communication Association},
  pages     = {5036--5040},
  doi       = {10.21437/Interspeech.2020-3015}
}

@inproceedings{stooke2024aligner,
  author    = {Adam Stooke and Rohit Prabhavalkar and Khe Chai Sim and Pedro Moreno Mengibar},
  title     = {Aligner-Encoders: Self-Attention Transformers Can Be Self-Transducers},
  booktitle = {Advances in Neural Information Processing Systems 37 (NeurIPS 2024)},
  year      = {2024}
}

@inproceedings{lee2021interctc,
  author    = {Jaesong Lee and Shinji Watanabe},
  title     = {Intermediate Loss Regularization for {CTC}-based Speech Recognition},
  booktitle = {2021 IEEE International Conference on Acoustics, Speech and Signal Processing (ICASSP)},
  pages     = {6224--6228},
  year      = {2021},
  doi       = {10.1109/ICASSP39728.2021.9414594}
}

@inproceedings{kim2017jointctcattn,
  author    = {Suyoun Kim and Takaaki Hori and Shinji Watanabe},
  title     = {Joint {CTC}-Attention based End-to-End Speech Recognition using Multi-task Learning},
  booktitle = {2017 IEEE International Conference on Acoustics, Speech and Signal Processing (ICASSP)},
  pages     = {4835--4839},
  year      = {2017},
  doi       = {10.1109/ICASSP.2017.7953075}
}

@inproceedings{sanabria2018slt,
  author    = {Ramon Sanabria and Florian Metze},
  title     = {Hierarchical Multi Task Learning With {CTC}},
  booktitle = {2018 IEEE Spoken Language Technology Workshop (SLT)},
  pages     = {485--490},
  year      = {2018},
  doi       = {10.1109/SLT.2018.8639530}
}

@inproceedings{higuchi2022icassp,
  author    = {Yosuke Higuchi and Keita Karube and Tetsuji Ogawa and Tetsunori Kobayashi},
  title     = {Hierarchical Conditional End-to-End {ASR} with {CTC} and Multi-Granular Subword Units},
  booktitle = {2022 IEEE International Conference on Acoustics, Speech and Signal Processing (ICASSP)},
  pages     = {7797--7801},
  year      = {2022},
  doi       = {10.1109/ICASSP43922.2022.9746580}
}

@inproceedings{kusunoki24_interspeech,
  title     = {{Hierarchical Multi-Task Learning with CTC and Recursive Operation}},
  author    = {Nahomi Kusunoki and Yosuke Higuchi and Tetsuji Ogawa and Tetsunori Kobayashi},
  year      = {2024},
  booktitle = {{Interspeech 2024}},
  pages     = {2855--2859},
  doi       = {10.21437/Interspeech.2024-1542}
}

@inproceedings{dong2020cif,
  author    = {Linhao Dong and Bo Xu},
  title     = {CIF: Continuous Integrate-And-Fire for End-To-End Speech Recognition},
  booktitle = {2020 IEEE International Conference on Acoustics, Speech and Signal Processing (ICASSP)},
  pages     = {6079--6083},
  year      = {2020},
  doi       = {10.1109/ICASSP40776.2020.9054250}
}

\end{document}